\documentclass[12pt,a4paper]{article}

\usepackage[margin=0.7in]{geometry}
\usepackage{multirow}

\usepackage[numbers]{natbib}
\usepackage{cite}

\usepackage{amsmath}
\usepackage{amssymb}

\usepackage[T1]{fontenc}
\usepackage[utf8]{inputenc}

\usepackage{lmodern}
\usepackage{microtype}

\usepackage{graphicx}
\graphicspath{{fig/}}

\usepackage{authblk}

\usepackage{xcolor}
\usepackage{hyperref}
\usepackage{orcidlink}

\newcommand{\ORCID}[1]{\orcidlink{#1}}

\providecommand{\keywords}[1]{%
  \par\noindent\textbf{Keywords: }#1%
}

\title{Integrating gene regulatory priors into Transformer attention with scTransformer for interpretable scRNA-seq analysis}

\author[1]{Mikele Milia\,\ORCID{0000-0002-2386-5475}}
\author[2,3]{Louis Fabrice Tshimanga\,\ORCID{0009-0002-1240-4830}}
\author[4]{Henning M\"uller\,\ORCID{0000-0001-6800-9878}}
\author[2,3,4]{Manfredo Atzori\,\ORCID{0000-0001-5397-2063}}
\author[1,5,6,$\ast$]{Barbara Di Camillo\,\ORCID{0000-0001-8415-4688}}

\affil[1]{Department of Information Engineering, University of Padova, Padova, Italy}
\affil[2]{Department of Neuroscience, University of Padova, Padova, Italy}
\affil[3]{Padova Neuroscience Center, Padova, Italy}
\affil[4]{Information Systems Institute, University of Applied Sciences Western Switzerland, HES-SO Valais, Sierre, Switzerland}
\affil[5]{Department of Comparative Biomedicine and Food Science, University of Padova, Padova, Italy}
\affil[6]{Padua Center for Network Medicine, University of Padova, Padova, Italy}
\affil[$\ast$]{Corresponding author: \href{mailto:barbara.dicamillo@unipd.it}{barbara.dicamillo@unipd.it}}

\date{}

\begin{document}

\maketitle

\begin{abstract}
\noindent\textbf{Motivation:} Transformer-based models are increasingly applied to large-scale single-cell transcriptomics, showing strong performance through self-supervised learning on millions of cells. However, most existing approaches treat genes as independent features, and largely ignore prior biological knowledge, which limits interpretability and robustness. In this paper, we explore whether explicitly incorporating gene regulatory information can improve both model performance and biological insight. \textbf{Results:} We present \textit{scTransformer}, the first Transformer-based approach that builds a priori knowledge of biological mechanisms into the model's attention patterns. By constraining information flow according to known regulatory structures, the model learns representations that are more biologically meaningful. We evaluate scTransformer on a disease-relevant single-nucleus RNA-seq dataset using supervised cell-type classification. Compared to standard Transformers, our approach improves classification accuracy, enhances separation of cell types in embedding space, and produces attention patterns consistent with known regulatory programs. Overall, our results demonstrate that embedding biological structure into Transformer models can enhance interpretability without sacrificing performance, offering a principled step toward biologically grounded foundation models for single-cell omics.\textbf{Contact:} \href{barbara.dicamillo@unipd.it}{barbara.dicamillo@unipd.it}
\end{abstract}



\section{Introduction}
Single-cell transcriptomics (scRNA-seq) enables the systematic characterization of cellular heterogeneity across development, homeostasis and disease. However, scRNA-seq data remain challenging to model: these data are high-dimensional and sparse, contain strong technical and biological confounders and often require analysis across cohorts, protocols and conditions. These challenges have motivated growing interest in representation learning methods that can use large, heterogeneous collections of single-cell profiles and provide features that transfer to downstream tasks. Transformer-based foundation models have emerged as a promising solution \citep{vaswani2017attention}. scGPT \citep{cui2024scgpt} and Geneformer, for instance, pretrain Transformer encoders on large corpora of single-cell expression profiles and transfer the learned representations to tasks such as cell-type annotation and perturbation modeling \citep{cui2024scgpt, theodoris2023transfer}. A shared strength of these models is scalability: by optimizing generic pretraining objectives on millions of cells, they can capture broadly useful representations. However, recent comparative evaluations reveal that performance and transferability vary substantially between settings and that the benefits over simpler baselines are not always consistent \citep{kedzierska2025zero}. A central limitation of existing single-cell Transformers is their reliance on \emph{unconstrained self-attention}: any gene token may attend to any other, and the model discovers dependencies solely from statistical co-variation. Although this design maximizes expressivity, it also expands the hypothesis space and makes attention patterns difficult to interpret mechanistically. In regimes that commonly arise in translational research, limited labeled data, shifts in cell-state composition across cohorts, and batch effects, unconstrained attention can overfit to dataset-specific correlations, yielding predictions that are neither robust nor biologically interpretable. In contrast, decades of regulatory genomics have produced curated transcription factor (TF) -- target gene (TG) resources \citep{garcia2019benchmark, muller2023expanding} that encode plausible causal structure in transcriptional control. Graph-based methods have exploited such priors for pathway activity inference and regulatory network reconstruction \citep{aibar2017scenic, holland2020transfer}, but they typically operate on fixed graphs and do not integrate directly with Transformer architectures. More recently, a few single-cell methods have used pathway-informed tokens or graph attention \citep{yang2022scbert, cui2022scformer} to bias learning towards known structures; however, no existing Transformer-based framework systematically integrates curated TF--TG networks into the attention mechanism itself.
Here we introduce \textit{scTransformer}, a biologically informed Transformer that embeds TF--TG priors \emph{inside} multi-head self-attention via hard masks that constrain which gene--gene interactions are admissible. Each attention head is prior-gated by a mask derived from a curated regulatory adjacency matrix, so that information flows only along curated, mechanistically grounded regulatory edges (and self-loops) while the model learns context-specific weighting over the admissible interactions. This architectural constraint provides a principled inductive bias that (i) improves classification accuracy, (ii) enhances robustness to noise and batch effects, (iii) increases sample efficiency in low-data regimes, and (iv) yields attention weights that are directly interpretable as regulatory programs, facilitating biological insight without post-hoc analysis. We evaluate scTransformer on the MSSM single-nucleus RNA-seq atlas \citep{fullard2025population}, comparing prior-gated models against unconstrained baselines, across training regimes on full and reduced data. Our experiments systematically assess each of the four hypothesized benefits: we quantify classification accuracy, measure reproducibility using cross-seed variance, visualize attention patterns relative to the regulatory network, and analyze performance degradation under data shrinkage. Our main contributions are: (i) a \emph{prior-gated attention} mechanism that converts curated TF--TG networks into hard attention masks, with variants encoding different assumptions about directionality and regulatory modules; (ii) a \emph{continuous sinusoidal encoding} of expression values that preserves quantitative information across the dynamic range of scRNA-seq measurements; (iii) a \emph{sparse batching and training pipeline} operating directly on compressed sparse row (CSR) matrices, enabling scalable training on large single-cell datasets; (iv) a formal set of \emph{module-level interpretability metrics} to quantify concentration and regulatory importance in attention maps; (v) a \emph{reproducible distributed-training protocol} with controlled data-partitioning, multi-seed evaluation, and explicit scaling diagnostics (see Supplementary Section~\ref{supp:sec:implementation}-\ref{supp:sec:ddp}).  Together, these elements outline a path for embedding biological structure into foundation-style models for single-cell omics, bridging flexible representation learning with mechanistic, biologically grounded interpretability.

\section{Methods}
\label{sec:methods}
\subsection{Problem formulation and notation}
We consider the task of learning cell representations by classifying single-cell RNA-sequencing (scRNA-seq) data, optionally guided by prior knowledge of transcriptional regulation.

\subsubsection{Inputs}
Let $\mathbf{X} \in \mathbb{R}^{N \times G}$ denote a single-cell expression matrix with $N$ cells and $G$ genes. Each row $\mathbf{x}_c$ is a sparse vector of normalized expression values, where $x_{c,g} \geq 0$ indicates the expression level of gene $g$ in cell $c$. Let $\mathcal{N}_c = \{g : x_{c,g} > 0\}$ denote the set of genes expressed in the cell $c$, with $L_c = |\mathcal{N}_c|$ usually ranging from hundreds to a few thousand. We additionally assume access to a curated transcription factor--target gene (TF--TG) network $\mathcal{G} = (\mathcal{V}, E)$, where $\mathcal{V}$ is the set of genes and $E \subseteq \mathcal{V} \times \mathcal{V}$ contains directed edges $(s, t)$ indicating that transcription factor $s$ regulates target gene $t$.

\subsubsection{Outputs}
The model produces a $d$-dimensional cell embedding $\mathbf{z}_c \in \mathbb{R}^d$ for each cell. For supervised tasks, this embedding is mapped to class predictions $\hat{\mathbf{y}}_c \in \mathbb{R}^K$ over $K$ cell types.

\subsubsection{Notation summary}
Throughout, we use: $G$ (total genes), $N$ (cells), $L$ (sequence length after padding), $d$ (embedding dimension), $H$ (attention heads), $d_k = d/H$ (per-head dimension), and $K$ (output classes).

\subsection{Model architecture}

scTransformer follows an encoder-only Transformer design with Pre-LN \citep{vaswani2017attention, xiong2020layer}, with token embeddings adapted to represent gene expressions.

\subsubsection{Gene token embedding.}
Each expressed gene $g$ in cell $c$ is mapped to a $d$-dimensional token by combining a gene identity embedding with an expression encoding:
\begin{equation}
  \mathbf{h}_{c,g}^{(0)} = \mathbf{e}^{\text{id}}_g + \mathbf{e}^{\text{expr}}_{c,g}.
\end{equation}
The gene identity embedding $\mathbf{e}^{\text{id}}_g \in \mathbb{R}^d$ is a trainable look-up indexed by gene identity (shared across all cells), with random initialization. The expression encoding $\mathbf{e}^{\text{expr}}_{c,g}$ uses a continuous sinusoidal scheme that generalizes positional encodings to continuous-valued inputs. Given the maximum expression range $x_{\max}$ and base $m = 2 x_{\max}$, we define the frequency coefficients $\omega_k = m^{-2k/d}$ for $k = 0, \ldots, \lfloor d/2 \rfloor - 1$, and encode each expression value as:
\begin{align}
  e^{\text{expr}}_{c,g}[2k] &= \sin(x_{c,g} \cdot \omega_k), \\
  e^{\text{expr}}_{c,g}[2k+1] &= \cos(x_{c,g} \cdot \omega_k).
\end{align}
This multi-scale representation ensures smoothness (similar expression values map to nearby embeddings) and boundedness regardless of input magnitude.

\subsubsection{Transformer encoder stack.}
The model comprises $N_{\text{layers}}$ stacked encoder layers. Each layer applies Pre-LN multi-head self-attention followed by a feed-forward network (FFN), both with residual connections:
\begin{align}
  \mathbf{h}' &= \mathbf{h}^{(\ell)} + \mathrm{MultiHead}(\mathrm{LN}(\mathbf{h}^{(\ell)}), M) \\
  \mathbf{h}^{(\ell+1)} &= \mathbf{h}' + \mathrm{FFN}(\mathrm{LN}(\mathbf{h}'))
\end{align}
where $M$ is an optional attention mask (detailed in Section~\ref{sec:prior-attention}). The feed-forward network uses a standard MLP with ReLU activation followed by a Gated Linear Unit (GLU) \citep{dauphin2017language} variant network \citep{shazeer2020glu}:
\begin{align}
  \mathrm{FFN}_{\text{MLP}}(\mathbf{x}) &= W_2 \sigma_{\text{MLP}}(W_1 \mathbf{x} + b_1) + b_2 \\
  \mathrm{FFN}_{\text{GLU}}(\mathbf{x}) &= W_2 (W_1 \mathbf{x} \odot \sigma_{\text{GLU}}(W_1' \mathbf{x})) + b_2
\end{align}
The hidden dimension is $2d$, with dropout applied after activation.

\subsubsection{Similarity-based pooling.}
\label{subsec:sim-pool}
To aggregate variable-length token sequences into a fixed-dimensional cell representation, we employ a similarity-based pooling mechanism. First, we compute a context vector as the (padding-aware) mean of token embeddings:
\begin{equation}
  \mathbf{q}_c = \frac{1}{|\mathcal{N}_c|} \sum_{g \in \mathcal{N}_c} \mathbf{h}_{c,g}^{(N_{\text{layers}})}.
\end{equation}
This context vector is the single-query in cross-attention over the token sequence:
\begin{equation}
  \mathbf{z}_c = \mathrm{MultiHead}(\mathrm{LN}(\mathbf{q}_c), \mathrm{LN}(\mathbf{H}_c), \mathrm{LN}(\mathbf{H}_c)),
\end{equation}
where $\mathbf{H}_c$ is the matrix of token embeddings for cell $c$. 
Padding and target gene tokens are masked, thus the learned pooling relies on transcription factor tokens, that are already interpolations of their respective target tokens.
This pooling learns to weight tokens according to their relevance to the cell-level representation. The resulting embedding is mapped to class logits via a linear projection.

\subsection{Biologically-informed attention}
\label{sec:prior-attention}

The key contribution of scTransformer is the integration of curated TF--TG regulatory knowledge from CollecTRI \citep{muller2023expanding} directly into the self-attention mechanism. Rather than allowing unconstrained attention where any gene may attend to any other, we construct a \emph{hard prior} that restricts information flow to biologically plausible interactions.

\subsubsection{Regulatory network preprocessing.}
Starting from the TF--TG network $\mathcal{G} = (\mathcal{V}, E)$, we filter to retain only edges where both the transcription factor (source gene) and the target gene are present in the dataset under analysis. To focus on well-characterized regulators, we optionally apply a target-count threshold $\tau$, keeping only TFs with more than $\tau$ unique targets:
\begin{equation}
  \mathcal{T} = \left\{ s : |\{t : (s,t) \in E\}| > \tau \right\}.
\end{equation}
For each TF $s \in \mathcal{T}$, let $\Gamma^{+}(s) = \{t : (s,t) \in E\}$ denote its target genes; for each target $t$, let $\Gamma^{-}(t) = \{s : (s,t) \in E\}$ denote its regulating TFs.

\subsubsection{Prior mask.}

\begin{figure}[!ht]
        \centering
        \includegraphics[width=0.4\linewidth]{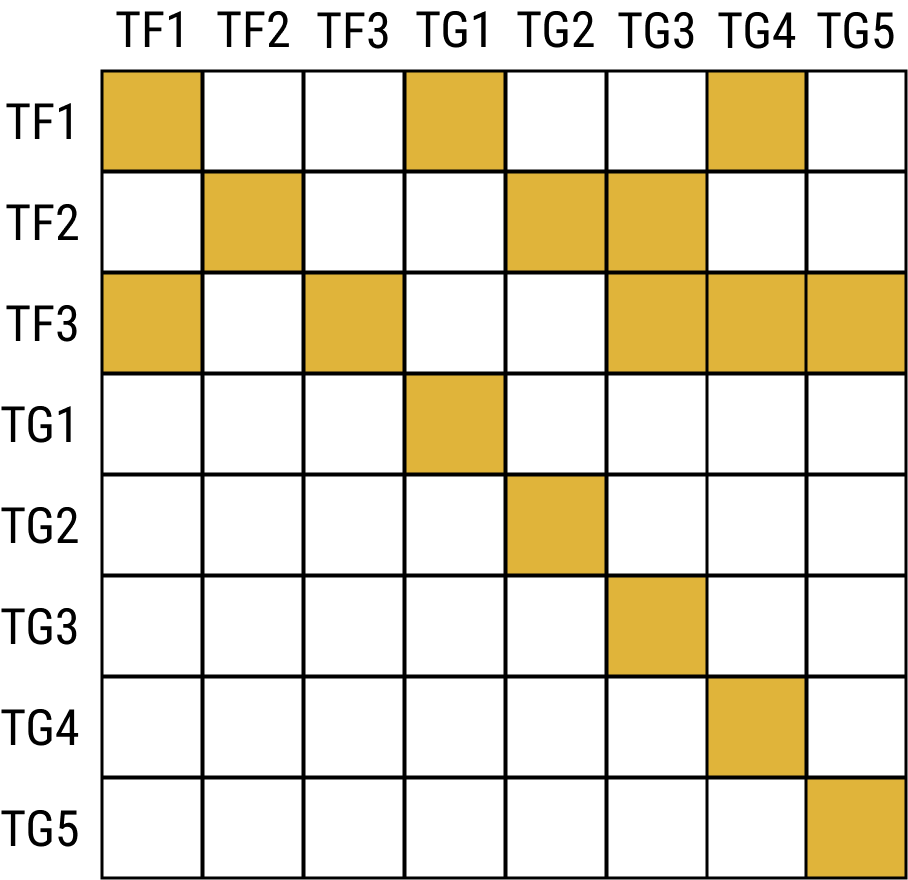}
        \caption{Illustrative example of directed TF→TG attention prior. Each yellow block denotes an allowed regulatory interaction from transcription factors (TFs) to their target genes (TGs), producing an asymmetric and highly sparse attention pattern. Empty regions correspond to biologically implausible connections that are masked out, guiding the model to focus on known regulatory pathways while preserving flexibility within the permitted structure.}
        \label{fig:asymmetric_prior}
    \end{figure}

We construct a boolean allow-matrix $M_{\text{allow}} \in \{0,1\}^{G \times G}$ over the full gene space, encoding different assumptions about regulatory information flow. Namely the encoding can be: 
\begin{itemize}

    \item[-] \textbf{unconstrained:} all gene–gene interactions are admissible ($M_{\text{allow}}(i,j)=1$ for all $i,j$).
    
    \item[-] \textbf{directed TF$\rightarrow$TG prior:} regulatory flow is enforced by allowing TF tokens $s \in \mathcal{T}$ to attend only to themselves and their targets ($M_{\text{allow}}(s,j)=1$ if $j \in \Gamma^{+}(s)\cup\{s\}$), while non-TF tokens are restricted to self-attention ($M_{\text{allow}}(i,j)=\mathbf{1}$ for $i\notin\mathcal{T}$). Self-loops are explicitly enforced for all tokens in this setting.
    
\end{itemize}
Fig.~\ref{fig:asymmetric_prior} illustrates this directed TF $\rightarrow$ TG prior, highlighting the resulting sparse, directed attention structure induced by the regulatory graph, no bidirectional masking variant is considered in this study. Efficient batch-level construction of the prior mask is described in Supplementary Section~\ref{supp:sec:implementation}.

\subsubsection{Prior-gated attention mechanism.}
The boolean allow-matrix is converted to an additive mask:
\begin{equation}
  \widetilde{M}_{\text{prior}}(i,j) =
  \begin{cases}
    0 & \text{if } M_{\text{allow}}(i,j) = 1, \\
    -\infty & \text{otherwise}.
  \end{cases}
\end{equation}
In each attention head $h$, this mask is added to the scaled dot-product scores before softmax:
\begin{equation}
  A_h = \mathrm{softmax}\left(\frac{Q_h K_h^\top}{\sqrt{d_k}} + \widetilde{M}_{\text{prior}} + M_{\text{pad}}\right),
\end{equation}
where $M_{\text{pad}}$ is an additional mask for padding tokens. The $-\infty$ entries yield exactly zero attention weights, guaranteeing that forbidden interactions do not contribute to the output. The prior attention mask, encoding structural constraints on permissible token interactions, is shared across all attention heads. Although each head learns distinct  $Q$/$K$/$V$ projections, they operate under the same masking constraints. For efficient batch computation, the mask is broadcast across heads and reshaped to match the attention score tensor, resulting in a final shape of $(B \cdot H, L, L)$ where $B$ is the batch size, $H$  the number of heads, and $L$ the sequence length. When using directed TF$\rightarrow$TG priors, we optionally restrict the pooling layer (discussed in Subsection~\ref{subsec:sim-pool}) to account only for TF tokens (genes with outgoing regulatory edges). This focuses the cell representation on transcriptional regulators rather than on all expressed genes.

\subsubsection{Regulatory-module interpretability metrics}
\begin{figure}[!ht]
    \centering
    \includegraphics[width=.8\linewidth]{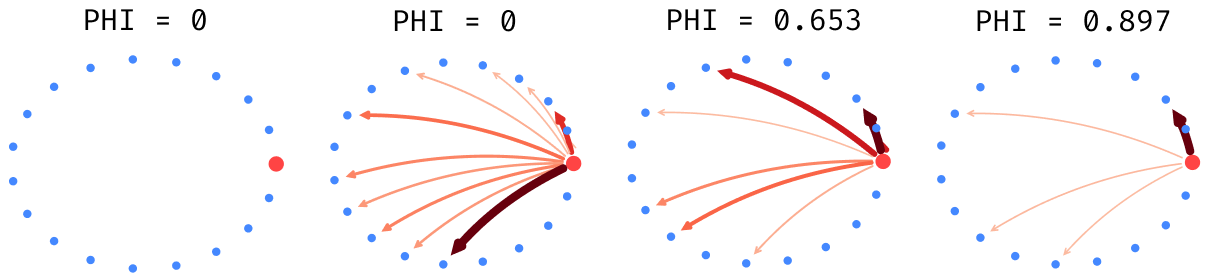}
    \caption{Conceptual illustration of $\phi$ score. The red node represents a transcription factor (TF), while blue nodes denote its potential target genes (TGs). From left to right, increasing concentration score reflects a transition from uniformly distributed (or absent) attention to progressively focused allocation on a smaller subset of targets. Edge color intensity encodes attention weight magnitude. Higher concentration corresponds to lower attention entropy and increased structural selectivity in TF–TG interactions.}
    \label{fig:focus_score_illustration}
\end{figure}

To assess the interpretability of regulatory modules, we quantify two complementary aspects of attention: (i) how selectively a transcription factor concentrates attention over its annotated targets, and (ii) the total amount of attention mass directed toward those targets. 
To quantify how each retained TF distributes attention over its annotated targets, we evaluate the distribution of attention weights from a given TF to its annotated targets and measure how concentrated this distribution is. Intuitively, if attention is spread uniformly across all targets, the module exhibits low structural selectivity; conversely, if attention is focused on a small subset of targets, the module displays a more structured and potentially biologically meaningful pattern. We capture this notion of selectivity using a normalized entropy-based concentration measure. 
we define compact entropy-based metrics on class-level attention maps.
Let $s \in \mathcal{T}$, where $\mathcal{T}$ is the filtered TF set defined above, and let $\Gamma^{+}(s)$ be its target genes. For each class $k$, layer $\ell$, and head $h$, we aggregate post-softmax cell-level attentions and compute a row-consistent class average:
\begin{equation}
A_{k,h}^{(\ell)}[i,j]
=
\frac{1}{n_{k,i}}
\sum_{c\in\mathcal C_k:\, i\in\mathcal N_c}
A_{c,h}^{(\ell)}[i,j],
\end{equation}
where $n_{k,i}=|\{c\in\mathcal C_k: i\in\mathcal N_c\}|$ is the number of class-$k$ cells in which query gene $i$ is present.

\[
w_i = A_{k,h}^{(\ell)}[s,\, t_i].
\]
For $|\Gamma^{+}(s)| \ge 2$, define
\begin{equation}
p_i =
\frac{w_i}{\sum_{t_j \in \Gamma^{+}(s)} w_j},
\qquad t_i \in \Gamma^{+}(s).
\end{equation}
\noindent The edge entropy is
\begin{align}
H &= - \sum_{t_i \in \Gamma^{+}(s)} p_i \log p_i, \\
H_{\mathrm{unif}} &= - \sum_{i=1}^{|\Gamma^{+}(s)|} \frac{1}{|\Gamma^{+}(s)|}\log\!\left(\frac{1}{|\Gamma^{+}(s)|}\right) = \log |\Gamma^{+}(s)|
\end{align}
\noindent with normalized entropy ratio
\begin{equation}
\eta = \frac{H}{H_{\mathrm{unif}}} \in [0,1].
\end{equation}
We define the concentration score
\begin{equation}
\phi = 1-\eta,
\end{equation}
Fig.~\ref{fig:focus_score_illustration} provides a conceptual visualization of the concentration score, illustrating how increasing values correspond to progressively more concentrated and structurally selective attention distributions, and where $\phi=0$ corresponds to a uniform outgoing pattern over annotated targets. To quantify the overall strength of the regulatory signal within a module, we sum the attention weights from the TF to its annotated targets. This term reflects how much total attention the model allocates to known regulatory edges, independently of how that attention is distributed. By combining these two components, structural concentration and total attention mass, we obtain a module-importance score that rewards both selective and substantial regulatory focus. In particular, module importance is defined as:
\begin{equation}
\mathcal{I}_{k,h}^{(\ell)}(s) = \phi \cdot \sum_{t_i \in \Gamma^{+}(s)} w_i.
\end{equation}

\noindent Using the last encoder layer ($\ell=L$) yields one score per module, class, and head:
\begin{equation}
T_{m,k,h}=\mathcal{I}_{k,h}^{(L)}(m),
\qquad
\mathbf{T}\in\mathbb{R}^{M\times C\times H},
\end{equation}
where $m$ indexes TF modules. For each $(k,h)$, we normalize over modules
$q_{m\mid k,h}=T_{m,k,h}/\sum_{m'}T_{m',k,h}$ and compute the module-distribution entropy
\begin{align}
H_{k,h}^{(M)} &=-\sum_{m=1}^{M} q_{m\mid k,h}\log q_{m\mid k,h}, \\
\eta_{k,h}^{(M)} &=\frac{H_{k,h}^{(M)}}{\log M},
\end{align}
with concentration $1-\eta_{k,h}^{(M)}$.
For degenerate cases ($|\Gamma^{+}(s)|<2$ or zero outgoing mass), the concentration is set to $0$.

\subsection{Implementation}

We used the MSSM PsychAD cohort described in~\citep{fullard2025population}, restricting the analysis to 17 well-represented subtypes. Gene features were limited to those present in the intersection between the dataset and the CollecTRI regulatory resource. TFs with fewer than 15 annotated targets were excluded to ensure sufficient regulatory connectivity. Full dataset preprocessing, quality-control criteria, and gene-space filtering steps are detailed in Supplementary Section~\ref{supp:sec:dataset}.
Samples were stratified into train/validation/test partitions (70/20/10). To evaluate scaling behavior, we progressively reduced the size of the training and validation sets while keeping the test set fixed. This protocol ensures controlled comparison across data regimes without altering the evaluation distribution.
Models were trained using AdamW ($\eta=10^{-4}$, $\lambda=10^{-5}$). The effective batch size was 128. Early stopping (patience 5, max 50 epochs) was applied based on validation loss. All hyperparameters and optimization settings were kept fixed across data regimes. Each experiment was repeated with 10 random seeds and results are reported as mean ± standard deviation. 
Mini-batches were constructed dynamically under a fixed token budget, with padding and masking applied as needed. Regulatory priors were indexed on-the-fly for each batch to ensure alignment between TFs and their targets. Reproducibility controls and distributed training configuration are reported in Supplementary Sections~\ref{supp:sec:implementation}--\ref{supp:sec:ddp}.

\section{Results}

We evaluate the impact of incorporating a regulatory prior across progressively larger training subsets (1\%, 25\%, 50\%, 75\%, 100\%), sampled with preserved class proportions while keeping validation and test sets fixed. This setup allows us to assess whether biologically informed attention provides advantages in low-data regimes and how these effects evolve with increasing sample size. We evaluate two factors: (i) feature-space restriction (full gene set vs. TF–TG-restricted genes) and (ii) the presence or absence of the directed TF→TG prior, across progressively reduced training subsets, with validation and test partitions held fixed. As shown in Table~\ref{tab:full_results}, constraining the model with a regulatory prior, rather than using fully unconstrained input interactions, is beneficial in low-data regimes and remains competitive at larger sample sizes. 

\begin{figure}[!ht]
    \centering
    \includegraphics[width=.8\linewidth]{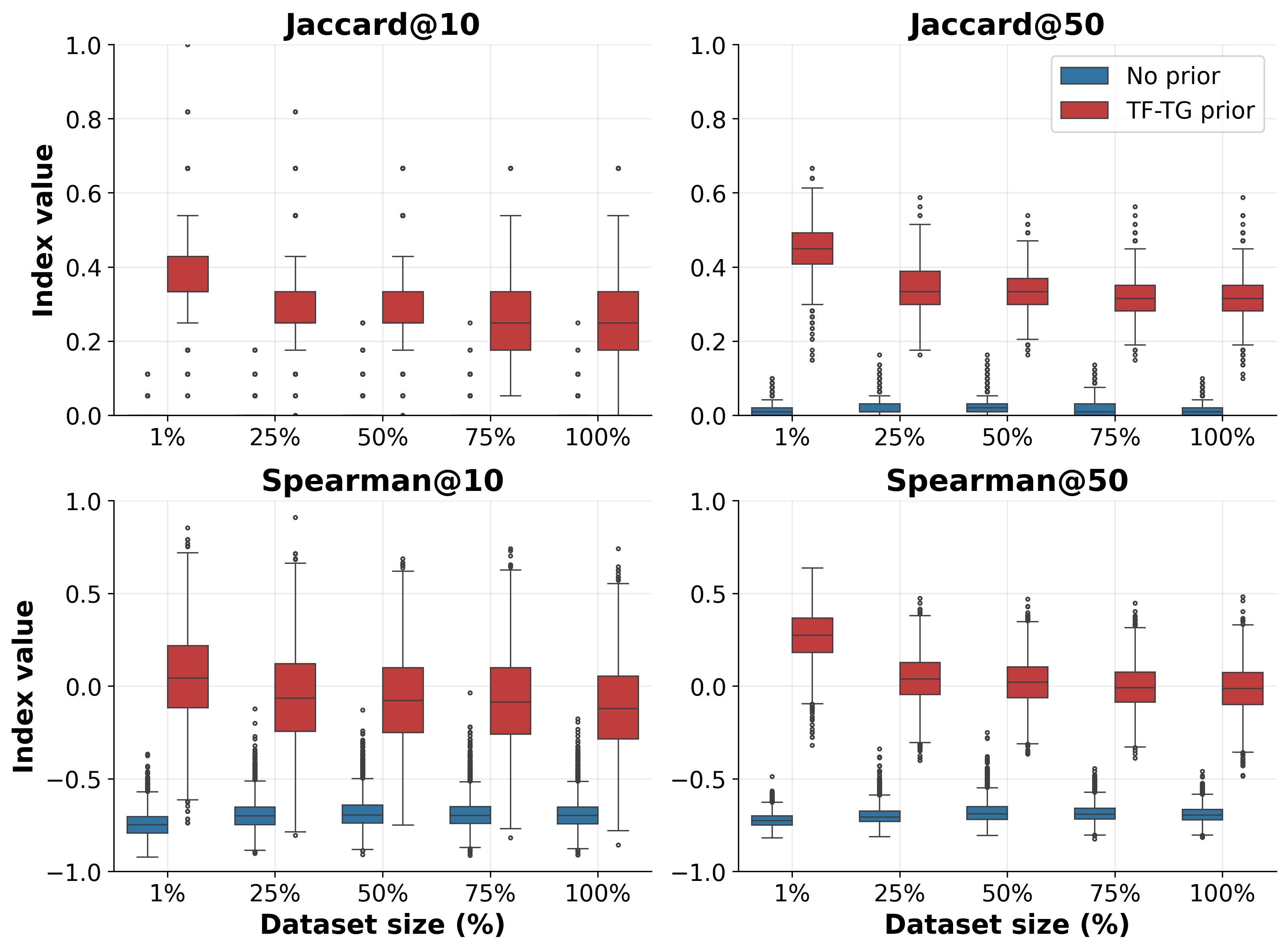}
    \caption{Run-to-run stability of top-$N$ selected genes across dataset sizes. Jaccard overlap (top) quantifies agreement in set membership, while Spearman correlation (bottom) captures consistency in gene ranking. Models trained without priors show near-zero agreement across runs, indicating that distinct gene combinations can support similar predictive performance. In contrast, prior-gated models exhibit consistently higher stability across both metrics, with the largest gains observed in low-data regimes but remaining detectable even at full data availability.}
    \label{fig:stability_vs_dataset}
\end{figure}

\begin{table}[!ht]
\centering
\caption{Accuracy across dataset sizes with and without the directed TF$\rightarrow$TG prior. Results are reported as mean $\pm$ std over 10 runs.}
\label{tab:full_results}
\renewcommand{\arraystretch}{1.15}
\setlength{\tabcolsep}{6pt}
\begin{tabular}{| l | cc |}
\hline
\multirow{2}{*}{Dataset} & \multicolumn{2}{c|}{Accuracy} \\
& No prior & TF$\rightarrow$TG prior \\
\hline

All genes (100\%)
 & 0.917 $\pm$ 0.012 & N/A \\

Only TF–TG (100\%)
 & 0.908 $\pm$ 0.008 & \textbf{0.911 $\pm$ 0.013} \\

Only TF–TG (75\%) 
 & \textbf{0.905 $\pm$ 0.009} & 0.903 $\pm$ 0.008 \\

Only TF–TG (50\%) 
 & 0.897 $\pm$ 0.008 & \textbf{0.899 $\pm$ 0.008} \\

Only TF–TG (25\%) 
 & 0.869 $\pm$ 0.012 & \textbf{0.878 $\pm$ 0.015} \\
Only TF–TG (1\%) 
 & 0.464 $\pm$ 0.059 & \textbf{0.486 $\pm$ 0.066} \\

\hline
\end{tabular}
\end{table}
Importantly, the 0.911 accuracy with TF$\rightarrow$TG priors is obtained under a substantially restricted interaction space (TF–TG genes only), yet remains close to the full-gene unconstrained baseline (0.917). At 1\% and 25\% of training data, prior-gated models achieve higher accuracy than the unconstrained baseline, indicating improved sample efficiency when training data are scarce. At 50\%--100\%, performance remains comparable across all data availability regimes, despite operating on a biologically restricted interaction space. Overall, these results suggest that selecting interactions \emph{a priori} can reduce the effective search space and stabilize learning. However, predictive performance alone does not capture structural consistency across runs. We therefore quantify cross-run reproducibility of the top-$N$ selected genes using Jaccard overlap and Spearman rank correlation (Fig.~\ref{fig:stability_vs_dataset}). Although models without priors achieve competitive accuracy, they show markedly lower stability, with different runs converging to distinct gene subsets that yield similar performance. This variability is consistent with the high-dimensional, sparse structure of scRNA-seq data, where multiple correlated feature subsets can support comparable predictions. This lack of reproducibility is particularly pronounced in the low-data regime, where the absence of inductive bias allows the model to rely on different subsets of correlated features. Although predictive performance remains comparable, the unconstrained model shows minimal cross-run agreement in selected genes. As shown in Fig.~\ref{fig:stability_vs_dataset}, Jaccard indices are close to zero and Spearman correlations are consistently negative, indicating that different runs select largely disjoint gene sets and rank them inconsistently. This instability causes the no-prior distributions to appear visually compressed relative to the prior-gated model. In contrast, the prior-gated model increases agreement across runs in both set membership (Jaccard index) and ranking consistency (Spearman correlation), with the largest differences observed for smaller $N$. These results indicate that incorporating the TF–TG prior reduces variability in selected feature subsets while maintaining comparable predictive performance. Across data regimes, models without priors achieve similar accuracy but rely on less stable gene combinations, whereas prior-gated models yield more consistent selections.

\subsection{TF regulatory modules exhibit increased focus under prior-gated attention}

Using the regulatory-module metrics defined in Section~\ref{sec:methods}, we compared attention focus and importance between prior-gated and unconstrained models. Directed TF$\rightarrow$TG masking consistently increased focus scores $\phi$, indicating that transcription factors allocate attention to smaller, more specific subsets of their annotated targets. This effect was most pronounced in intermediate encoder layers and was accompanied by higher module importance for biologically coherent TF programs, suggesting that regulatory priors promote structured and interpretable attention patterns. This effect is particularly relevant in sparse transcriptomic settings, where weakly constrained models may otherwise distribute attention across interchangeable feature sets. To assess whether this increased focus translates into biologically meaningful representations, we examined TF modules at full data availability (100\%), where predictive performance is comparable between models. In this regime, differences cannot be attributed to sample efficiency, but rather to how the model organizes regulatory information. We observe that prior-gated attention enables TF modules to selectively activate target programs associated with specific cellular subtypes. In contrast, unconstrained models tend to distribute attention more diffusely, resulting in weaker or inconsistent module activation. As illustrated in Fig.~\ref{fig:no_prior_top5_tf_modules}, prior-guided modules align with distinct cell-type-specific regulatory patterns, whereas the no-prior model fails to recover comparable structure despite achieving similar classification accuracy. Applying the same module-selection procedure to the unconstrained model yields more diffuse and less subtype-localized patterns (Supplementary Fig.~\ref{fig:no_prior_top5_tf_modules}), indicating that the regulatory prior primarily affects the organization of transcriptional programs rather than predictive performance.

\begin{figure}[!ht]
    \centering
    \includegraphics[width=\linewidth]{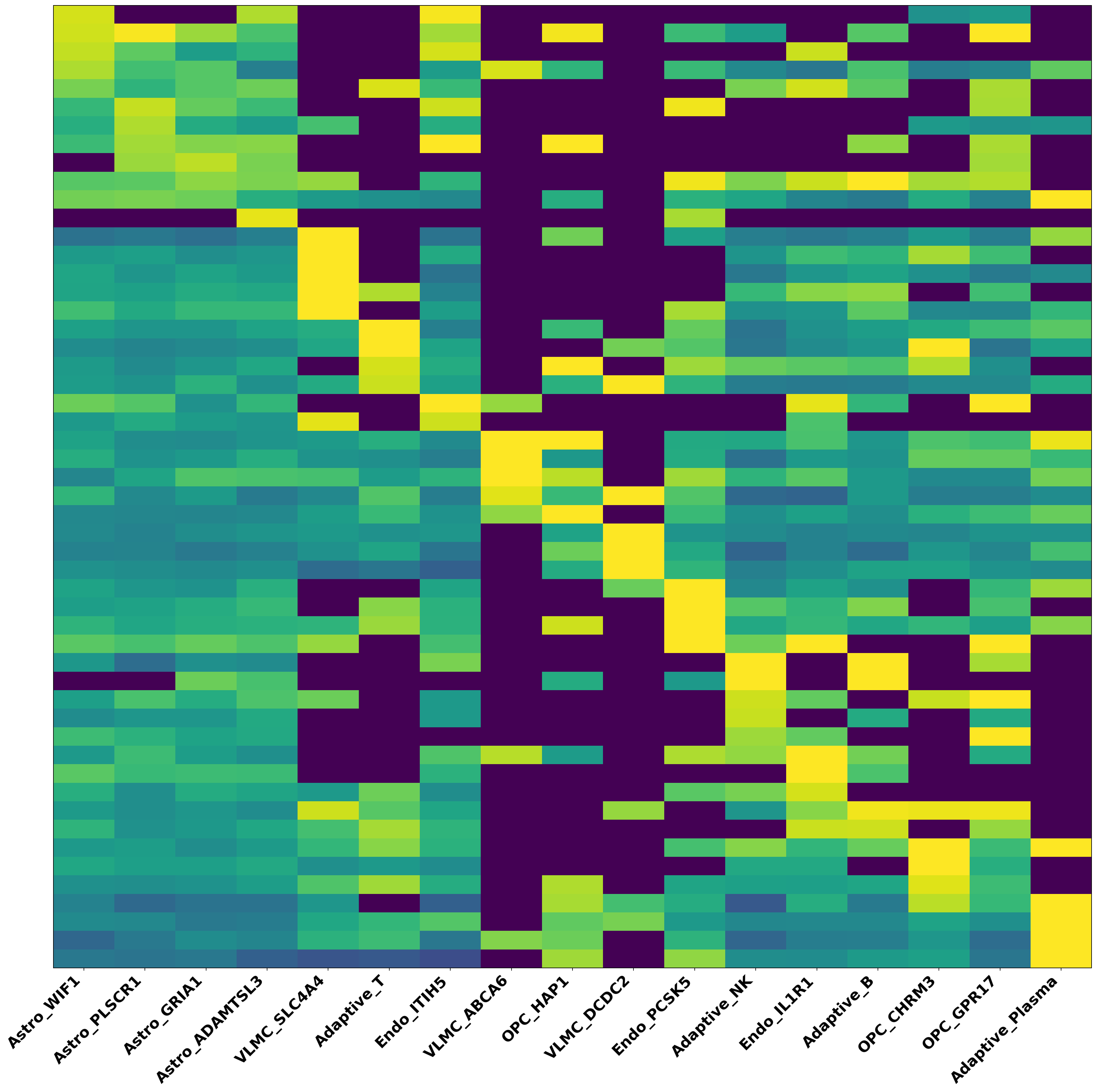}
    \caption{Union of top-5 TF modules across cell types. Rows report TF modules selected among the highest-scoring components for each subtype and merged into a non-redundant set. The heatmap shows the distribution of focus scores across cell types, illustrating how regulatory modules tend to localize within specific subtypes while remaining partially shared across related lineages.}
    \label{fig:with_prior_top5_tf_modules}
\end{figure}

These findings indicate that the regulatory prior does not only bias feature selection, but promotes functionally coherent transcriptional programs. In settings where performance alone is insufficient to distinguish models, the prior resolves ambiguity by steering attention toward biologically grounded interaction patterns that are consistent with known TF--TG relationships.
To provide a broader view across cell types, we constructed a union of the top-5 transcriptional modules identified for each subtype. Specifically, for each cell type, we selected its highest ranked modules and progressively merged them into a non-redundant set (Fig.~\ref{fig:with_prior_top5_tf_modules}). The resulting heatmap highlights how high-focus TF modules tend to concentrate within specific subtypes rather than being uniformly shared across cell types.
Although overlap remains, as expected in partially shared regulatory programs, dominant signals exhibit localized patterns, consistent with subtype-aligned regulatory organization. In this case, prior-gated attention is associated with a more localized concentration of high-scoring TFs in the highlighted subtype, while focus remains comparatively distributed across other cell types. Although this example is not intended as definitive evidence on its own, similar patterns are observed across multiple TF modules, suggesting that the prior can encourage subtype-aligned regulatory activation without enforcing rigid structure. 

Since accuracy is matched in this regime, these structural differences cannot be explained by improved predictive fit; they instead reflect how the prior constrains the hypothesis space, reducing non-identifiable solutions that rely on interchangeable gene subsets.
Inspection of the most recurrent transcription factors within the top-ranked modules reveals enrichment for regulators associated with lineage specification and cell-state plasticity, including \textit{GRHL3}, \textit{DLX2}, and \textit{NANOG}. Notably, these modules exhibit subtype-localized patterns: identity-associated regulators such as \textit{GRHL3} and \textit{CDX1} are enriched in endothelial-like and VLMC populations, whereas developmental and plasticity-linked factors such as \textit{DLX2} and \textit{NANOG} are more prominent in OPC and adaptive subtypes. A detailed biological interpretation of these subtype-associated regulatory programs is provided in Supplementary Section~\ref{supp:sec:bio_modules}.

\section{Discussion}

In this work, we introduce a biologically informed deep learning framework for single-cell classification, the first Transformer that integrates regulatory structure directly into the attention mechanism. At the architectural level, we propose a prior-gated attention mechanism that transforms curated TF–TG networks into hard structural attention masks. Unlike post-hoc regularization strategies or soft penalties, this design embeds regulatory structure directly into the computation of attention weights, constraining the space of admissible interactions during training. The framework retains flexibility while imposing biologically meaningful inductive bias. This architectural constraint reduces the structural non-identifiability inherent in high-dimensional learning problems, guiding the model toward coherent regulatory programs rather than arbitrary gene combinations that achieve similar predictive performance. In unconstrained models, the sparsity and dimensionality of scRNA-seq data render the classification problem inherently underdetermined. Multiple distinct gene subsets can explain the same phenotypic signal, leading to solutions that are predictive yet unstable across training runs. While such solutions may optimize accuracy, they lack structural consistency and limit biological interpretability. By incorporating TF–TG priors directly into the attention mechanism, we restrict the hypothesis space to biologically plausible interaction patterns.

As a result, the learned solutions exhibit greater reproducibility in feature selection and more stable transcription factor module activation across runs, improving the reliability of downstream biological interpretation. At the representation level, we introduce a continuous sinusoidal encoding of gene expression values that preserves quantitative information across the full dynamic range of scRNA-seq measurements. Rather than discretizing or thresholding expression levels, this encoding provides a smooth, bounded, and multi-scale representation of expression magnitudes. This design allows the model to exploit subtle quantitative differences while maintaining numerical stability, ensuring that regulatory attention mechanisms operate on information-rich representations rather than coarse approximations. From a computational perspective, we develop a sparse batching and training pipeline operating directly on CSR matrices, avoiding dense tensor expansion and minimizing padding overhead. This implementation enables scalable training on large single-cell datasets while preserving architectural flexibility and regulatory constraints. 

Empirically, we provide a systematic evaluation across data regimes demonstrating that regulatory priors confer multiple benefits. Prior-gated models achieve predictive accuracy that is competitive with unconstrained baselines, while exhibiting enhanced robustness to noise and batch effects, improved reproducibility of inferred regulatory modules, and increased sample efficiency in low-data settings. Notably, the advantages of the prior are most pronounced when sample size is limited, where model variance and feature-selection instability are otherwise amplified. As dataset size increases, predictive performance remains comparable between constrained and unconstrained models, indicating that regulatory priors enhance robustness without restricting flexibility when sufficient data are available. Collectively, these findings show that regulatory priors enhance not only the stability of learned representations but also the reliability of downstream biological interpretation. By steering attention toward biologically plausible interaction patterns, the prior improves robustness to noise and batch effects, increases reproducibility of inferred regulatory modules, and preserves predictive performance across data regimes. In this sense, incorporating regulatory structure does more than regularize the model: it improves the epistemic reliability of the conclusions that can be drawn from it, providing a principled framework for interpretable deep learning in regulatory genomics.

While scTransformer demonstrates the utility of embedding regulatory priors into Transformer attention, the following concepts suggest directions for future work. First, the current implementation relies on \emph{hard} binary masks, fully blocking interactions absent from the curated network. Although this guarantees strict adherence to prior knowledge, it prevents the model from overriding the prior when data strongly support novel interactions. Soft or learnable priors, where disallowed edges receive a finite penalty, could enable a controlled trade-off between biological bias and data-driven discovery, potentially allowing to increase current knowledge within the domain. Second, our evaluation is restricted to supervised cell-type classification. Transformer-based foundation models benefit substantially from self-supervised pretraining on large unlabeled corpora; extending scTransformer to masked-gene reconstruction with prior-gated attention may improve generalization and enable transfer learning across datasets. Third, applying regulatory priors requires restricting the gene set to annotated TF--TG pairs, excluding unannotated genes that may still carry biologically relevant signal. Hybrid architectures that combine prior-gated attention for annotated genes with unconstrained attention for others could address this limitation. Finally, experiments were conducted on a single RNA-seq dataset, leaving generalization across tissues, diseases, and protocols untested. In addition, constructing batch-specific attention masks introduces modest computational overhead, which may become more pronounced at larger scales; sparse attention implementations could help mitigate this cost.

\section{Conclusion}

In this work, we evaluated the role of TF--TG regulatory priors in guiding attention-based models for cell-type classification across varying data regimes.
Although prior-gated models achieve predictive performance comparable to unconstrained baselines at full data availability, they provide improved sample efficiency in low-data settings and promote more stable feature selection across runs. Beyond predictive metrics, we showed that regulatory priors influence how the model organizes transcriptional signals. In the absence of priors, models can converge to distinct gene combinations that solve the classification task equally well, leading to solutions that are predictive but structurally inconsistent. By constraining the interaction space, prior-gated attention encourages the emergence of more coherent TF modules and subtype-aligned regulatory patterns without sacrificing flexibility. Taken together, these findings indicate that the primary benefit of incorporating regulatory knowledge lies not in improving prediction alone, but in enhancing the consistency and interpretability of the learned representations. In high-dimensional sparse data, such as single-cell transcriptomics, this constraint helps resolve otherwise ambiguous solutions that remain predictive but biologically unstable. In settings where multiple solutions can explain the data, regulatory priors help resolve this ambiguity by steering attention toward biologically grounded interaction patterns.

\section*{Availability and implementation}
The datasets analyzed in this study are publicly available from the CZ CELLxGENE Discover data portal (Chan Zuckerberg Initiative) and described in \citep{fullard2025population}. All data were accessed in accordance with the repository’s terms of use. 

\section*{Fundings}

This work is partially supported by the HEREDITARY Project, as part of the European Union's Horizon Europe research and innovation programme under grant agreement No GA 101137074.

\section*{Acknowledgements}

The authors gratefully acknowledge the support of the CONVECS project (CUP C97H23001700002), funded by Veneto Region in the PR Veneto FESR 2021-2027 framework, Action 1.1.2 Sub B.

\bibliographystyle{unsrtnat}
\small
\bibliography{reference}


\newpage


\newcounter{suppsection}
\newcounter{suppsubsection}[suppsection]
\newcounter{suppsubsubsection}[suppsubsection]

\renewcommand{\thesuppsection}{S\arabic{suppsection}}
\renewcommand{\thesuppsubsection}{S\arabic{suppsection}.\arabic{suppsubsection}}
\renewcommand{\thesuppsubsubsection}{S\arabic{suppsection}.\arabic{suppsubsection}.\arabic{suppsubsubsection}}

\renewcommand{\theHsuppsection}{supp.section.\arabic{suppsection}}
\renewcommand{\theHsuppsubsection}{supp.section.\arabic{suppsection}.\arabic{suppsubsection}}
\renewcommand{\theHsuppsubsubsection}{supp.section.\arabic{suppsection}.\arabic{suppsubsection}.\arabic{suppsubsubsection}}

\newcommand{\suppsection}[1]{%
  \refstepcounter{suppsection}%
  \section*{\thesuppsection\quad #1}%
  \addcontentsline{toc}{section}{\thesuppsection\quad #1}%
}

\newcommand{\suppsubsection}[1]{%
  \refstepcounter{suppsubsection}%
  \subsection*{\thesuppsubsection\quad #1}%
  \addcontentsline{toc}{subsection}{\thesuppsubsection\quad #1}%
}

\newcommand{\suppsubsubsection}[1]{%
  \refstepcounter{suppsubsubsection}%
  \subsubsection*{\thesuppsubsubsection\quad #1}%
  \addcontentsline{toc}{subsubsection}{\thesuppsubsubsection\quad #1}%
}

\renewcommand{\thefigure}{S\arabic{figure}}
\renewcommand{\thetable}{S\arabic{table}}
\renewcommand{\theequation}{S\arabic{equation}}

\setcounter{figure}{0}
\setcounter{table}{0}
\setcounter{equation}{0}

\section*{Supplementary Material}
\addcontentsline{toc}{section}{Supplementary Material}

\noindent This Supplementary Material reports the operational details underlying the experiments in the main text. It includes precise specifications of data preprocessing, batching construction, distributed training setup, and reproducibility controls. All conceptual design choices and methodological rationale are described in the main manuscript; the present sections provide the parameters necessary for exact replication. Dataset and preprocessing details are reported in Section~\ref{supp:sec:dataset}, implementation and infrastructure details in Section~\ref{supp:sec:implementation}, distributed training analyses in Section~\ref{supp:sec:ddp}, unconstrained-attention comparison in Section~\ref{supp:sec:no_prior}, biological module interpretation in Section~\ref{supp:sec:bio_modules}, and embedding-level comparison in Section~\ref{supp:sec:umap}.

\suppsection{MSSM snRNA-seq dataset (PsychAD cohort)}
\label{supp:sec:dataset}
We utilized the Mount Sinai NIH Brain Bank and Tissue Repository (MSSM) subset of the PsychAD cohort described in \citep{fullard2025population}. The full PsychAD cohort comprises 1,494 post-mortem human donors, of which 1,042 were sourced from MSSM, making MSSM the largest contributing site within the consortium. Tissue samples were obtained from the dorsolateral prefrontal cortex (DLPFC) and profiled using single-nucleus RNA sequencing (snRNA-seq). Experiment-specific data selection and split-level filtering are detailed in Section~\ref{supp:subsec:dataset_usage}.
This section is organized into library preparation (Section~\ref{supp:subsec:dataset_library}), quality-control criteria (Section~\ref{supp:subsec:dataset_qc}), experiment-level data usage (Section~\ref{supp:subsec:dataset_usage}), and biological subtype characterization (Section~\ref{supp:subsec:dataset_bio}).

\suppsubsection{Library preparation and sequencing}
\label{supp:subsec:dataset_library}
Nuclei were processed using the 10x Genomics Chromium 3' v3.1 platform. 
Sequencing reads were aligned to the human reference genome (hg38) using STARsolo. 
Genotype-based demultiplexing was performed using cellSNP and vireo.

\suppsubsection{Quality control and filtering (as reported in the original study)}
\label{supp:subsec:dataset_qc}
Across the PsychAD cohort, nuclei were retained if they satisfied the following criteria:
\begin{itemize}
    \item[-] total UMI counts between 1,500 and 110,000;
    \item[-] detected genes between 1,100 and 12,500; 
    \item[-]  mitochondrial read fraction $<$ 5\%.
\end{itemize}
Ambient RNA contamination was corrected using CellBender, and putative doublets were removed using Scrublet.
Genes expressed in at least 0.05\% of nuclei were retained.
Donor samples with fewer than 50 nuclei were excluded.

\suppsubsection{Data usage}
\label{supp:subsec:dataset_usage}

Within scTransformer, data filtering was applied as follows:
\begin{itemize}

    \item[-] \textbf{Cell-type subset filter.} We retained only nuclei whose subtype annotation starts with one of the following prefixes: \textit{Adaptive\_}, \textit{Astro\_}, \textit{Endo\_}, \textit{OPC\_}, \textit{VLMC\_}. This restriction was introduced to focus the benchmark on a biologically coherent subset of MSSM cell populations with sufficient sample size per class, reducing extreme class imbalance and improving comparability across dataset-size regimes. This yields 17 subtype classes used in all experiments.
      
    \item[-] \textbf{Gene-space filter guided by TF--TG prior.} Starting from the CollecTRI TF--TG resource (\citep{muller2023expanding}), we first retained TFs with strictly more than 15 unique targets, then built the union of source/target genes and intersected it with the dataset feature (gene) space. In the MSSM setting, this produced 6,094 candidate genes from the interaction table, of which 5,558 are present in the splits and used as model input (same gene order enforced across train/val/test).

\end{itemize}

\noindent For the 100\% MSSM split, the filtering pipeline gives: train \(2{,}898{,}317 \rightarrow 417{,}703\) nuclei, validation \(828{,}090 \rightarrow 119{,}371\), and test \(414{,}046 \rightarrow 59{,}721\) (about \(85.6\%\) reduction in each split). The reduction in \emph{nuclei count} is attributable to the cell-type subset restriction only. The TF--TG gene-space restriction does not remove nuclei; it changes the feature space (genes) used by the model. QC filters (UMI/genes/mitochondrial fraction, doublets, ambient RNA) were applied upstream by the original PsychAD processing and are therefore not an additional reduction step in our scTransformer split-level filtering. The same filtering strategy was applied consistently across the 1\%, 25\%, 50\%, 75\%, and 100\% data regimes.

\suppsubsection{Biological characterization of MSSM-derived cellular subtypes}
\label{supp:subsec:dataset_bio}
Subtype-specific summaries are reported in Sections~\ref{supp:ssubsec:dataset_bio_astro}, \ref{supp:ssubsec:dataset_bio_opc}, \ref{supp:ssubsec:dataset_bio_endo}, \ref{supp:ssubsec:dataset_bio_vlmc}, and~\ref{supp:ssubsec:dataset_bio_adaptive}.

The single-nucleus RNA-seq data used in this study were derived from the MSSM subset of the PsychAD cohort, profiled from the dorsolateral prefrontal cortex (DLPFC), a region critically involved in executive function and highly vulnerable to neuropsychiatric and neurodegenerative pathology. To ensure both biological coherence and statistical robustness, we restricted the analysis to nuclei belonging to five major lineage-level compartments:

\begin{itemize}
\item[-] Astrocytic
\item[-] Endothelial
\item[-] Oligodendrocyte precursor cells (OPC)
\item[-] Vascular and leptomeningeal cells (VLMC)
\item[-] Adaptive / immune-associated populations
\end{itemize}

Within these compartments, a total of 17 transcriptionally resolved subtypes were retained based on the prefix-based annotation provided in the original PsychAD atlas. This selection balances two key constraints: preservation of lineage diversity relevant to neurovascular and glial regulatory dynamics, and avoidance of extreme class imbalance that could confound supervised learning.

\suppsubsubsection{Astrocytic subtypes}
\label{supp:ssubsec:dataset_bio_astro}
Astrocytes were represented by multiple transcriptionally distinct states, including \textit{Astro\_WIF1}, \textit{Astro\_PLSCR1}, \textit{Astro\_GRIA1}, and \textit{Astro\_ADAMTSL3}. These subtypes capture functional heterogeneity spanning synaptic support, extracellular matrix modulation, and signaling responsiveness, reflecting gradients of metabolic support, inflammatory tone, and synaptic coupling within the dorsolateral prefrontal cortex (DLPFC).

\suppsubsubsection{OPC lineage}
\label{supp:ssubsec:dataset_bio_opc}
The OPC compartment included \textit{OPC\_GPR17}, \textit{OPC\_HAP1}, and \textit{OPC\_CHRM3}. These represent maturation- and signaling-divergent precursor states rather than a homogeneous progenitor pool, capturing neuromodulatory responsiveness and intracellular specialization relevant to lineage plasticity.

\suppsubsubsection{Endothelial populations}
\label{supp:ssubsec:dataset_bio_endo}
Endothelial nuclei were subdivided into \textit{Endo\_IL1R1}, \textit{Endo\_ITIH5}, and \textit{Endo\_PCSK5}, reflecting inflammatory sensing, extracellular matrix interaction, and proteolytic processing programs. Inclusion of endothelial diversity is particularly relevant in regulatory-learning frameworks due to their role in barrier-associated transcriptional signaling.

\suppsubsubsection{VLMC populations}
\label{supp:ssubsec:dataset_bio_vlmc}
VLMC subtypes included \textit{VLMC\_ABCA6}, \textit{VLMC\_DCDC2}, and \textit{VLMC\_SLC4A4}, capturing functional specialization in lipid transport, cytoskeletal organization, and ion homeostasis, respectively. These cells contribute structural and extracellular interaction programs relevant to transcriptional regulation.

\suppsubsubsection{Adaptive populations}
\label{supp:ssubsec:dataset_bio_adaptive}
Adaptive-like nuclei comprised \textit{Adaptive\_B}, \textit{Adaptive\_T}, \textit{Adaptive\_NK}, and \textit{Adaptive\_Plasma}. Although not classical parenchymal brain cells, these populations capture immune surveillance and adaptive signaling states within the neurovascular niche.

Importantly, these 17 subtypes do not represent isolated biological entities but form a partially overlapping regulatory landscape. Some transcriptional programs are lineage-specific, while others reflect shared regulatory backbones across related populations. This hierarchical structure provides a biologically structured yet computationally tractable setting for evaluating how regulatory priors influence subtype-level organization of transcriptional programs.

\suppsection{Implementation and infrastructure details}
\label{supp:sec:implementation}
This Section includes sparse representation (Section~\ref{supp:subsec:impl_sparse}), on-the-fly prior indexing (Section~\ref{supp:subsec:impl_prior}), partitioning protocol (Section~\ref{supp:subsec:impl_partition}), and reproducibility policy (Section~\ref{supp:subsec:impl_repro}).

\suppsubsection{Sparse data representation}
\label{supp:subsec:impl_sparse}

Gene expression matrices are stored in compressed sparse row (CSR) format in the single-cell data object. The full $N \times G$ matrix is never materialized as a dense tensor during training. Instead, row-wise access is used to extract only non-zero genes for each cell, which are converted into variable-length token sequences prior to padding.

Dense representations are created only at the mini-batch level after padding, with sequence length $L$ determined by the dynamic batching policy (described in Section~\ref{supp:sec:ddp}). No global dense expansion over the full gene space is performed. This design reduces host-to-device transfer volume, limits peak memory usage, and avoids redundant intermediate dense tensors, which is particularly important in high-dimensional sparse transcriptomic data.

\suppsubsection{On-the-fly prior indexing}
\label{supp:subsec:impl_prior}

The global TF--TG adjacency derived from CollecTRI is stored once as a sparse index structure mapping each transcription factor (TF) to its list of target gene indices. A full $G \times G$ dense allow-mask is not constructed at runtime. During mini-batch construction, given the batch-specific gene index set (after padding to length $L$), the prior mask is generated by intersecting the batch gene indices with the precomputed global adjacency lists. Operationally, this results in a batch attention mask of shape $(L, L)$, without allocating a dense $G \times G$ structure. This strategy preserves exact alignment between TF tokens and their admissible targets while avoiding redundant storage and repeated preprocessing. The prior remains globally defined but is indexed dynamically at the batch level, ensuring consistency across epochs and data regimes.

\suppsubsection{Data partitioning protocol}
\label{supp:subsec:impl_partition}

The initial train/validation/test split (70/20/10), stratified by subtype label, is computed once and stored. All data-scaling regimes (1\%, 25\%, 50\%, 75\%, 100\%) reuse the same test partition. For reduced-data experiments, only the training and validation splits are subsampled, with preserved class proportions. The test split remains fixed across all regimes, and no cross-regime reshuffling is performed. This protocol guarantees that performance differences across dataset sizes are attributable solely to variation in training data availability, rather than to changes in evaluation distribution. It also ensures direct comparability of metrics across scaling experiments.

\suppsubsection{Determinism and reproducibility policy}
\label{supp:subsec:impl_repro}

All experiments use fixed random seeds controlling (i) data subsampling, (ii) model parameter initialization, and (iii) stochastic states in the numerical libraries used for training. Hyperparameters are kept identical across data regimes, and no regime-specific retuning is performed. Optimization settings, architectural parameters, and regularization coefficients remain constant for all experiments. Distributed training is executed with one process per GPU and synchronized gradient updates (see Section~\ref{supp:sec:ddp}). Exact bitwise determinism at the GPU-kernel level is not enforced; instead, cross-run variability is quantified explicitly by repeating each experiment with 10 independent seeds and reporting mean $\pm$ standard deviation. This policy prioritizes reproducibility at the statistical level while maintaining computational efficiency in distributed training.

\suppsection{Distributed training and computational speedup}
\label{supp:sec:ddp}
This section covers distributed setup (Section~\ref{supp:subsec:ddp_setup}), software/runtime reproducibility (Section~\ref{supp:subsec:ddp_runtime}), dynamic sparse batching (Section~\ref{supp:subsec:ddp_batching}), length bucketing and prior construction (Section~\ref{supp:subsec:ddp_bucketing}), memory behavior (Section~\ref{supp:subsec:ddp_memory}), speedup definitions (Section~\ref{supp:ssubsec:ddp_metrics}), and empirical log-based results (Section~\ref{supp:ssubsec:ddp_logs}).

\suppsubsection{Distributed data parallel setup}
\label{supp:subsec:ddp_setup}
Training is executed with a standard data-parallel distributed strategy, one process per GPU. Each process receives a distinct data shard, while gradients are synchronized at each optimization step through collective communication. In our production runs we used NVIDIA H100 80GB GPUs; the largest regimes (25--100\%) were trained with \(p=4\) GPUs, while the 1\% regime was trained with \(p=1\) GPU.

Across different \(p\), the token budget is kept constant \emph{per rank} (same sampler limits on each GPU), not constant globally. Therefore, the effective global token budget per optimization step increases approximately linearly with \(p\), and reported scaling should be interpreted as weak-scaling-oriented rather than strict strong scaling.

This choice is intentional: under strict strong scaling with a fixed global token budget, increasing $p$ would force progressively smaller per-rank batches/tokens, which in practice amplifies communication overhead (all-reduce latency dominance) and can reduce hardware efficiency. We therefore prioritize a stable per-rank workload and memory profile to obtain representative multi-GPU training behavior for this sparse, variable-length setting.

\suppsubsection{Software/runtime reproducibility details.}
\label{supp:subsec:ddp_runtime}
For the DDP scaling benchmark, training was run in a CUDA-enabled environment with distributed GPU communication support. Node-level logs confirm a consistent GPU driver/runtime stack across runs. Mixed precision was enabled in FP16 mode, and optimizer updates were performed at every batch (i.e., no additional gradient-accumulation steps at runtime). Optimizer hyperparameters follow the training configuration: Adam with learning rate \(10^{-4}\), weight decay \(10^{-5}\), betas \((0.9,0.999)\), and epsilon \(10^{-8}\).

\suppsubsection{Dynamic sparse batching with token budget.}
\label{supp:subsec:ddp_batching}
Mini-batches are dynamically constructed (not fixed-size) with epoch-wise greedy packing under explicit memory-oriented constraints: maximum token budget per batch (\(100{,}000\) tokens), minimum/maximum batch size (\(64/128\) in training), and maximum padding ratio (\(0.3\)). Operationally, if a candidate batch has size \(b\) and current maximum sequence length \(L_{\max}\), it is accepted only if:
\begin{equation}
b \cdot L_{\max} \le T_{\max}.
\end{equation}
with \(T_{\max}=100{,}000\). This policy limits peak memory usage while keeping high device utilization on sparse, variable-length inputs. In distributed mode, the global batch plan is built and then sharded across ranks; in our training setup it is rebuilt every epoch, so both batch composition and steps/epoch can vary slightly across epochs.

\suppsubsection{Length bucketing and batch-level prior construction}
\label{supp:subsec:ddp_bucketing}

Cells vary widely in the number of expressed genes. To minimize computational and memory overhead associated with padding variable-length sequences, cells are grouped (bucketed) according to their number of non-zero tokens, so that sequences of similar length are processed together within the same mini-batch. Padding is applied only up to the maximum sequence length present in the batch, thereby reducing the proportion of padding tokens and improving computational efficiency. Padding uses a dedicated index with zero expression, and key-padding masks are propagated through all attention layers to prevent padded positions from contributing to attention scores or value aggregation. Within each bucket, sampling is performed to ensure that each batch contains cells from at least half of the available classes, preventing homogeneous batches and stabilizing training dynamics.

Because each cell expresses only a subset of genes, attention priors are constructed on-the-fly by indexing into the global prior adjacency rather than materializing a dense gene-by-gene mask. For batch indices $\mathbf{I} \in \mathbb{N}^{B \times L_{\max}}$, the batch-specific prior mask is obtained as
\begin{equation}
M_{\text{batch}}[b,i,j] = \tilde{M}_{\text{prior}}[\mathbf{I}[b,i], \mathbf{I}[b,j]].
\end{equation}
Padding positions are masked, while diagonal entries remain unmasked to guarantee valid self-attention and preserve identity mappings.

\suppsubsection{Memory tuning and per-rank memory behavior.}
\label{supp:subsec:ddp_memory}
Memory stability is enforced primarily through token-budgeted dynamic batching rather than through a fixed sample count per batch. We tuned token budget, batch-size limits, and padding-ratio constraints to keep peak activation memory within the H100 80GB envelope while preserving throughput. Under DDP, each rank executes the same model with the same per-rank token budget; therefore, the per-rank memory footprint is expected to remain approximately constant with increasing $p$, since the model replica and per-rank token budget are unchanged. In particular, rank 0 does not require systematically higher model-activation memory than other ranks; observed differences are mainly transient system/runtime effects (allocator state, data-loading timing, communication buffers), not a structural increase with \(p\).

\suppsubsubsection{Speedup metrics.}
\label{supp:ssubsec:ddp_metrics}

Let \(T_p\) denote the post-warmup wall-clock time per epoch measured with \(p\) GPUs 
in the dedicated DDP benchmark (\(p \in \{1,2,4,8\}\)), under identical data split, 
optimizer state, and training protocol. The single-GPU baseline \(T_1\) is directly 
measured (not extrapolated) and serves as the reference for all reported values in 
Table~\ref{tab:ddp_speedup}.

\paragraph{Epoch-time speedup and efficiency.}
We first report classical epoch-level speedup:
\begin{equation}
S_p = \frac{T_1}{T_p}, 
\qquad 
E_p = \frac{S_p}{p},
\end{equation}
where \(S_p\) measures the reduction in time-to-epoch and 
\(E_p \in (0,1]\) quantifies parallel efficiency relative to ideal linear scaling.

\paragraph{Scaling regime.}
Because the per-rank token budget is kept constant as \(p\) increases, 
the total effective tokens processed per global step scale proportionally with \(p\). 
This corresponds to a \emph{weak-scaling-oriented} regime rather than strict strong scaling. 
Consequently, \(T_p\) captures both:

\begin{itemize}
\item \textbf{systems scaling effects}: computation/communication overlap, 
all-reduce latency, bandwidth contention, and synchronization overhead;
\item \textbf{workload variation}: small cross-\(p\) differences in effective work per epoch 
induced by dynamic sparse rebatching and packing efficiency.
\end{itemize}

Under this regime, ideal scaling would satisfy:
\[
T_p^{\text{ideal}} \approx T_1,
\]
up to communication overhead, since per-rank work is held constant.

\paragraph{Workload-normalized throughput speedup.}
To decouple systems scaling from workload drift, 
we additionally report a workload-normalized speedup based on global-step throughput:
\begin{equation}
S^{\text{work}}_p 
=
\frac{\left(N_{\text{steps,global},p} / T_p\right)}
     {\left(N_{\text{steps,global},1} / T_1\right)},
\qquad
E^{\text{work}}_p = \frac{S^{\text{work}}_p}{p},
\end{equation}
where \(N_{\text{steps,global},p}\) is the number of executed global optimization 
steps in one epoch at scale \(p\).

This metric normalizes for small variations in the effective number of global steps 
per epoch caused by dynamic sparse rebatching, thus isolating the scaling behavior 
of the distributed training system from packing-induced workload fluctuations.

\begin{table}[!h]
\centering
\caption{DDP scaling benchmark on the MSSM 50\% split (\(p \in \{1,2,4,8\}\)). Each value is computed from 3 independent runs; for each run we first average over post-warmup epochs, then aggregate across runs as mean \(\pm\) std. In addition to classical speedup/efficiency, we report workload-normalized metrics using global steps per epoch.}
\label{tab:ddp_speedup}
\begin{tabular}{rrrrrr}
\hline
\(p\) & \(\bar{T}_p\) (s/epoch) & \(N_{\text{steps,global}}\) & \(S_p\) & \(E_p\) & \(S^{\text{work}}_p\) \\
\hline
1 & \(100.76 \pm 0.67\) & \(2357\) & 1.000 & 1.000 & 1.000 \\
2 & \(75.10 \pm 0.25\)  & \(2392\) & 1.342 & 0.671 & 1.362 \\
4 & \(61.12 \pm 0.46\)  & \(2450\) & 1.649 & 0.412 & 1.713 \\
8 & \(53.57 \pm 0.06\)  & \(2511\) & 1.881 & 0.235 & 2.004 \\
\hline
\end{tabular}
\end{table}

While \(S_p\) reflects the user-facing metric of time-to-epoch, 
\(S^{\text{work}}_p\) more closely approximates hardware efficiency 
under equivalent effective work. Reporting both provides a clearer 
separation between algorithmic workload drift and distributed systems scalability.

\suppsubsubsection{Empirical results from training logs.}
\label{supp:ssubsec:ddp_logs}
To quantify DDP scaling under controlled conditions, we performed a dedicated benchmark on the MSSM 50\% split using \(p \in \{1,2,4,8\}\) GPUs (same model/configuration, same data split, and same training protocol across configurations). This benchmark is the primary source for reporting parallel speedup and efficiency. In this benchmark, \(T_1\) is directly measured (not estimated), so \(S_p\) and \(S^{\text{work}}_p\) are computed from observed runs. As expected in this weak-scaling-oriented setup, wall-clock speedup by epoch is sub-linear (\(S_8=1.881\)), while workload-normalized speedup is slightly higher (\(S^{\text{work}}_8=2.004\)) because \(N_{\text{steps,global}}\) increases with \(p\) (\(\sim 2357 \rightarrow 2511\), \(\approx 6.5\%\)). This drift is driven by dynamic repacking under fixed per-rank token budget (with length and padding constraints, plus epoch-wise rebuild), which changes global packing efficiency as sharding changes with \(p\). In this benchmark, workload normalization is therefore step-based; token-level totals per epoch were not explicitly logged, so residual token-level workload drift cannot be fully excluded. We did not run dedicated profiler traces to decompose per-step time into compute vs communication components, so the observed efficiency drop at higher \(p\) should be interpreted as an aggregate systems effect rather than an isolated communication-only effect.
Memory telemetry was sampled from rank-0 logs only (not averaged across ranks) using runtime counters reported during training: 

\begin{itemize}
    \item[-] \texttt{Mem} (device memory in use from the logger),
    \item[-] \texttt{Alloc} (allocated device memory),
    \item[-] \texttt{Reserved} (reserved device memory),
    \item[-] \texttt{MaxAlloc} (peak allocated device memory). 
\end{itemize}

Under this telemetry, peak rank-0 \texttt{Mem} remains stable across configurations, consistent with comparable per-rank memory under fixed per-rank token budget.
Warmup overhead across scales is reported in Table~\ref{tab:ddp_warmup}, and rank-0 memory peaks are summarized in Table~\ref{tab:ddp_memory}.

\begin{table}[!h]
\centering
\caption{Warmup overhead in the DDP benchmark (MSSM 50\%). Epoch 1 is compared against the post-warmup epoch-time mean pooled across runs.}
\label{tab:ddp_warmup}
\begin{tabular}{rrrr}
\hline
\(p\) & Epoch 1 (s) & Post-warmup mean (s) & Overhead \\
\hline
1 & 103.66 & 100.80 & \(+2.84\%\) \\
2 & 77.18  & 75.09  & \(+2.78\%\) \\
4 & 64.17  & 61.18  & \(+4.89\%\) \\
8 & 56.97  & 53.57  & \(+6.35\%\) \\
\hline
\end{tabular}
\end{table}

\begin{table}[!h]
\centering
\caption{Rank-0 memory peaks (GB) in the DDP benchmark, aggregated across 3 runs as mean \(\pm\) std.}
\label{tab:ddp_memory}
\begin{tabular}{rrrrr}
\hline
\(p\) & Mem peak & Alloc peak & Reserved peak & MaxAlloc peak \\
\hline
1 & \(25.63 \pm 0.10\) & \(7.05 \pm 0.00\) & \(24.25 \pm 0.09\) & \(21.04 \pm 0.00\) \\
2 & \(25.87 \pm 0.49\) & \(7.05 \pm 0.00\) & \(24.21 \pm 0.50\) & \(20.86 \pm 0.00\) \\
4 & \(25.80 \pm 0.51\) & \(7.05 \pm 0.00\) & \(24.08 \pm 0.51\) & \(20.55 \pm 0.00\) \\
8 & \(24.76 \pm 0.18\) & \(7.05 \pm 0.00\) & \(22.77 \pm 0.18\) & \(19.82 \pm 0.00\) \\
\hline
\end{tabular}
\end{table}

\suppsection{Comparison with unconstrained attention}
\label{supp:sec:no_prior}

\begin{figure}[!ht]
    \centering
    \includegraphics[width=.8\linewidth]{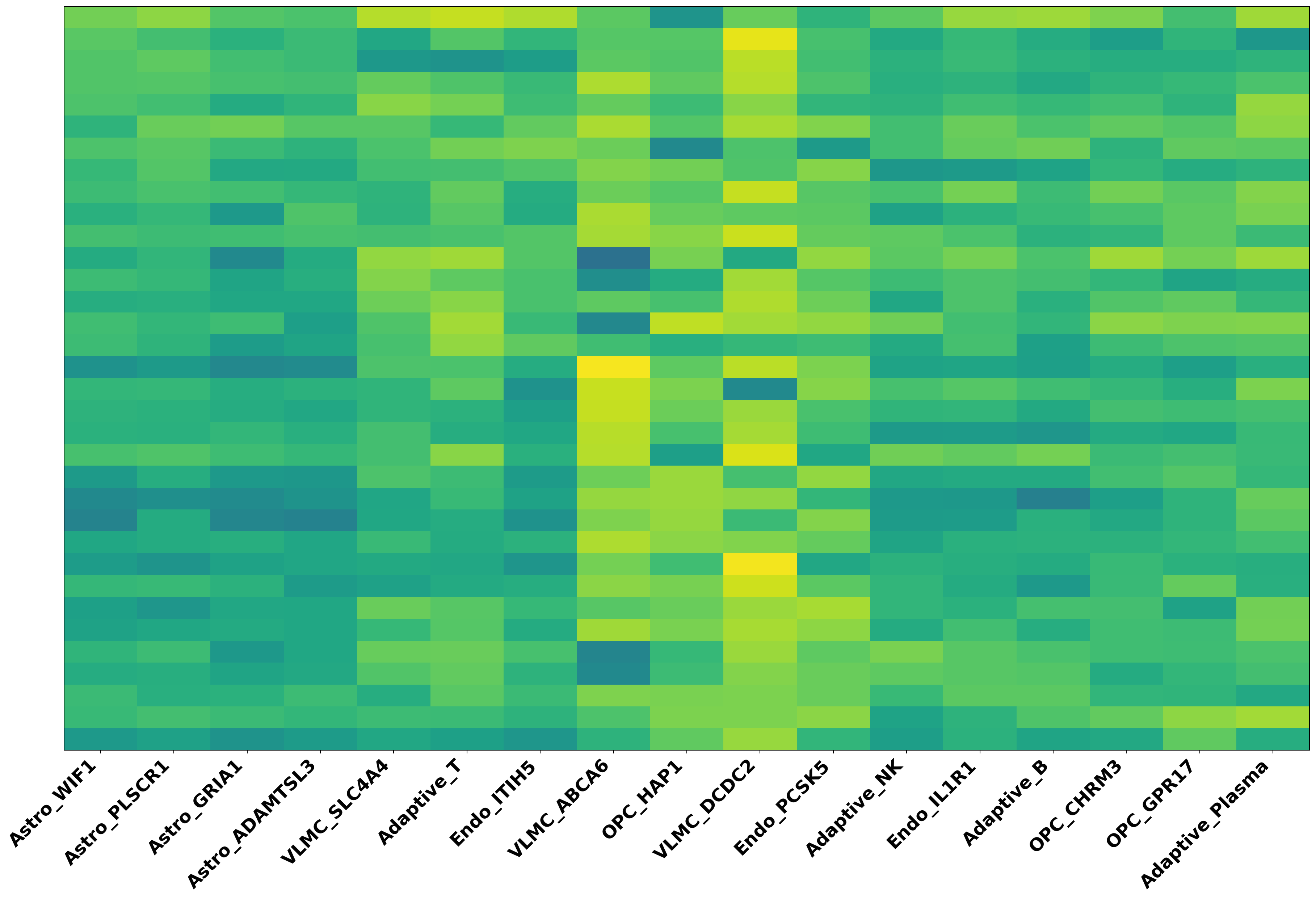}
    \caption{ Union of top-5 TF modules across cell types in the unconstrained model. Modules were ranked using the same entropy-weighted importance score (Eq.~19–20) used in the main analysis. Compared to the prior-gated model (Fig.~\ref{fig:with_prior_top5_tf_modules}), the unconstrained model exhibits broader cross-subtype activation and weaker subtype localization of high-focus TF modules, consistent with a more diffuse organization of regulatory signals in the absence of structural constraints.}    \label{fig:no_prior_top5_tf_modules}
\end{figure}

To determine whether the subtype-localized regulatory structure observed in Fig.~\ref{fig:with_prior_top5_tf_modules} arises specifically from the TF–TG prior, we repeated the same module-selection procedure using the unconstrained model at full data availability (100\%). Module ranking was performed using the identical entropy-weighted importance score (Eq.~19–20), ensuring that differences reflect architectural constraints rather than post-hoc scoring choices. Despite comparable predictive performance between models (Table~\ref{tab:full_results}), the unconstrained architecture yields qualitatively different regulatory organization. As shown in Fig.~\ref{fig:no_prior_top5_tf_modules}, top-ranked TF modules exhibit broader cross-subtype activation and reduced localization within specific cell types. Instead of concentrating attention within subtype-aligned programs, high-scoring modules tend to distribute their influence across multiple subtypes, resulting in more diffuse regulatory patterns.

This contrast indicates that the TF–TG prior primarily affects how transcriptional signals are structured rather than how well they support classification. In the absence of prior constraints, multiple gene combinations can explain subtype identity equally well, allowing the model to rely on interchangeable regulatory signals. The prior-gated model, in contrast, restricts the admissible interaction space and promotes more coherent subtype-aligned module activation. 

Taken together, these results support the interpretation that the regulatory prior does not mainly enhance predictive accuracy at full data availability, but instead improves the structural identifiability of transcriptional programs by encouraging localized and functionally consistent regulatory patterns. A complementary embedding-level comparison is provided in Section~\ref{supp:sec:umap}.

\suppsection{Biological characterization of subtype-associated regulatory modules}
\label{supp:sec:bio_modules}

To assess whether the transcription factor (TF) modules highlighted in Fig.~4 reflect biologically meaningful regulatory programs rather than dataset-specific statistical artefacts, we examined the union of the top-ranked modules across cell subtypes. Across cell types, the most recurrent TFs include \textit{GRHL3}, \textit{DLX2}, \textit{NANOG}, \textit{POU1F1}, and \textit{NFE2}. Importantly, these regulators are not canonical housekeeping controllers, but are instead associated with lineage specification, differentiation, and cell-state plasticity.

Several of these TFs are linked to regulatory programs consistent with the identity of specific cell classes represented in the dataset. Modules containing \textit{GRHL3} and \textit{CDX1}, both associated with epithelial lineage specification and barrier-associated transcriptional programs, are preferentially enriched in endothelial-like and VLMC populations, which are structurally specialized and interface with the extracellular environment. Similarly, modules involving \textit{FOXL2}, a regulator of fate stabilization, are observed in astrocytic subtypes, consistent with roles in maintenance of differentiated cellular identity.

Other modules highlight regulatory programs related to developmental progression and lineage maturation. \textit{DLX2}-associated modules are predominantly observed within OPC and adaptive-like populations, in line with its role in developmental patterning and lineage commitment, while modules involving \textit{MYOG} and \textit{BHLHA15} are enriched in VLMC and endothelial-associated subtypes, suggesting transcriptional programs linked to structural specialization and maturation.

In parallel, TFs such as \textit{NANOG} and \textit{NFE2} localize to modules enriched in adaptive and OPC subtypes. Given the progenitor-like characteristics of OPC populations and the functional responsiveness of adaptive classes, the presence of these regulators is consistent with transcriptional programs associated with state plasticity and adaptive responses rather than fixed lineage identity.

Notably, individual TF modules show subtype-preferential enrichment without being strictly subtype-exclusive. Several regulators recur across related populations, indicating a partially shared regulatory backbone. This pattern suggests a hierarchical organization in which some TFs contribute to broadly shared regulatory programs across multiple cell types, while others exhibit stronger subtype-localized enrichment consistent with specialized identity or functional state. Such partial overlap aligns with biological expectations, as related cell populations often share upstream regulatory architecture while diverging at downstream transcriptional programs.

Overall, the top-ranked modules are enriched for TFs implicated in lineage commitment, differentiation, and adaptive cellular states rather than ubiquitous transcriptional activity. Their distribution across cell types combines localized enrichment with partial sharing across related populations, supporting the view that prior-guided attention promotes the emergence of biologically structured regulatory programs aligned with cell-type-specific transcriptional organization.

\suppsection{Embedding-level organization with and without regulatory prior}
\label{supp:sec:umap}

\begin{figure}[!ht]
    \centering
    \includegraphics[width=0.7\textwidth]{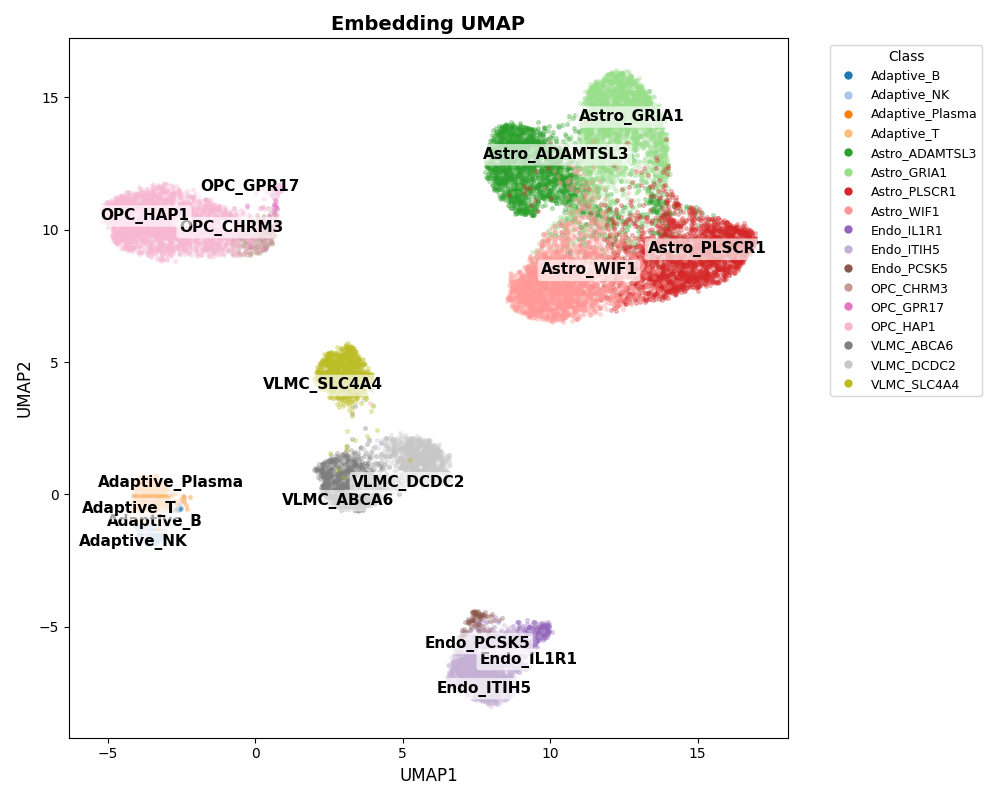}
    \caption{UMAP visualization of cell embeddings learned without the regulatory prior. Major lineage groups, including astrocytic, OPC, endothelial, VLMC, and adaptive populations, are clearly separated, indicating that the model captures broad biological structure even in the absence of prior constraints.}
    \label{fig:supp_umap_no_prior}
\end{figure}

\begin{figure}[!ht]
    \centering
    \includegraphics[width=0.7\textwidth]{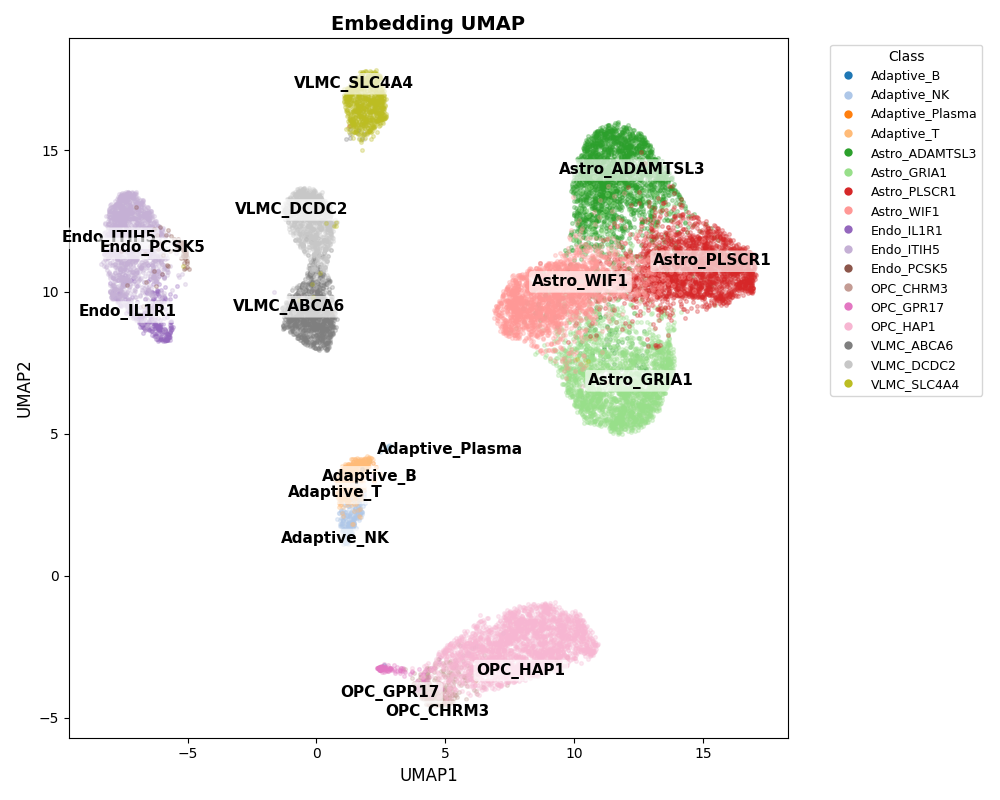}
    \caption{UMAP visualization of cell embeddings learned with the TF--TG regulatory prior. Lineage-level separation remains preserved, while subtype organization appears more localized, particularly within astrocytic and VLMC populations. This suggests that incorporating regulatory structure refines subtype-level coherence without disrupting global lineage relationships.}
    \label{fig:supp_umap_with_prior}
\end{figure}

To assess whether the regulatory prior influences the global organization of the learned representations, we visualized cell embeddings using UMAP for models trained with and without the TF--TG constraint at full data availability (100\% training set; Figs.~\ref{fig:supp_umap_no_prior} and~\ref{fig:supp_umap_with_prior}). In both cases, major lineage groups, including astrocytic, OPC, endothelial, VLMC, and adaptive populations, remain clearly separable, indicating that the prior does not artificially impose lineage structure in the high-data regime.

However, qualitative differences emerge in the internal organization of related subtypes. In the absence of the prior (Fig.~\ref{fig:supp_umap_no_prior}), subtype boundaries tend to appear more diffuse, with local mixing between closely related populations. In contrast, embeddings learned under prior-guided attention (Fig.~\ref{fig:supp_umap_with_prior}) display tighter subtype localization while preserving proximity among biologically related classes.

Notably, astrocytic and VLMC populations exhibit more structured internal gradients under the prior-constrained model, consistent with partially shared regulatory programs coupled with subtype specialization. Conversely, adaptive and endothelial subtypes remain spatially proximal in both settings, suggesting that the prior enhances subtype-level coherence without enforcing artificial separation.

Taken together, these observations indicate that incorporating regulatory structure does not alter the broad lineage-level organization of the embedding space, but refines subtype-level structure in a manner consistent with biologically interpretable transcriptional variation. This behavior aligns with the subtype-localized regulatory organization highlighted in Fig.~\ref{fig:with_prior_top5_tf_modules}.

\end{document}